**Photoluminescence spectra of an *n*-doped (Cd,Mn)Te quantum well: an exemplary evidence for the anisotropy-induced valence-band mixing**

A.V. Koudinov[1,2], C. Kehl[3], G. Astakhov[1,3], J. Geurts[3], T. Wojtowicz[4], G. Karczewski[4]

[1]A.F. Ioffe Physico-Technical Institute of RAS, 194021 St.-Petersburg, Russia

[2]Spin Optics Laboratory, St.-Petersburg State University, 198504 St.-Petersburg, Russia

[3]Physikalisches Institut, Universität Würzburg, 97074 Würzburg, Germany

[4]Institute of Physics, Polish Academy of Sciences, 02-668 Warsaw, Poland

Photoluminescence spectra of a (001)-$Cd_{0.99}Mn_{0.01}Te$ quantum well were taken with linear-polarization resolution and using an in-plane magnetic field. Because the quantum well contained a two-dimensional electron gas, the spectra consisted of several features. Since the quantum well layer was formed by a diluted magnetic semiconductor, the spectra showed pronounced polarization-dependent transformations when the in-plane magnetic field was applied. In the magnetic field, a 90-degrees rotation of the sample about the surface normal axis resulted in a clearly different spectrum, meaning that the nominally equivalent $[110]$ and $[1\bar{1}0]$ directions in the sample are not equivalent in fact. But, remarkably, the additional 90-degrees rotations of both the polarizer and the analyzer restored the initial spectrum. This combined invariance regarding simultaneous 90-degrees rotation of the sample and reversal of the polarization configuration was known earlier for spin-flip Raman spectra only. Our present observations are interpreted in terms of the mixing of valence subbands leading to the pseudo-isotropic *g*-factor of the ground-state holes.

**1. Introduction.**

Since the early 1990s, optical spectroscopy began to reveal the fact that epitaxial quantum well- and quantum dot heterostructures grown of diamond-like semiconductors frequently possess lower symmetry than it follows from simple geometric considerations involving ideal lattices.[1,2] As one of principal consequences, there arises the mixing of heavy- and light-hole subbands of the complex valence band $\Gamma_{15}$.[3,4] In particular, it results in the degeneracy removal of the doublet of "bright" exciton states $|\pm1\rangle$, which was confirmed by direct measurements of the optical spectra at a single-particle level,[5,6] in addition to numerous indirect experimental evidences. Apart from the electron-hole interaction within the exciton, the heavy-light hole mixing manifests itself in the interaction of holes with the external magnetic field $B$. This latter consequence shows up even in the absence of the fine structure of the spectrum, e.g., in case of the optical emission from trion states.[7]

In an ideal structure grown along the [001] direction, the $B$-linear (Zeeman) spin splitting of the hole states should be practically absent; in reality, however, such splitting shows up both in quantum wells (QWs)[8,9] and in single quantum dots (QDs) belonging to various heterostructure families (e.g., in CdSe,[7,10,11] CdTe,[12,13] InGaAs,[14] InAs,[15,16] GaAs,[17] etc.). This splitting, as well as specific features of the relevant optical transitions involving the split levels, originates from the anisotropic mixing perturbation in the lateral plane. The role and the nature of such perturbation has been analyzed in detail at a phenomenological and at a microscopic level,[9,13,18,19,20] and also experimentally. For example, the authors of Ref.[13] came to the conclusion that for CdSe quantum dots, the main source of the anisotropy comes from the in-plane strain through the Bir-Pikus Hamiltonian. Contrary to that, the authors of Refs.[17,21] concluded a dominating role of the QD shape. Whatever the case, the symmetry of the light-emitting state turns out to be reduced – down to $C_2$ or $C_{2v}$ at least, leading to the similar effect on the optical transitions.

The physics of valence-band mixing and pseudo-isotropic *g*-factor of holes is widely recognized in the field of the optics of single QDs where optical transitions involving separate spin sublevels can be resolved in the spectral domain and analyzed individually. For QWs, where the distinct transitions are buried within the broadened spectral lines, only indirect experimental evidences for the same physical mechanism were obtained (e.g., rather delicate measurements of the linear polarization degree of the photoluminescence (PL)).[8,9] However, a favorable model system can allow direct access to the impact of the mixing on the PL spectra. The details of this impact can convincingly demonstrate the symmetry of the optical transitions driven by the mixing, and can show that this physics works safely for quite ordinary (regarding symmetry and technology) QWs.

In the present paper, we describe transformations of the PL spectra of a (Cd,Mn)Te quantum well, subject to in-plane magnetic fields. We show how the polarization configuration of the PL experiment combines with the orientation of the in-plane field **B** to form a particular image of the PL spectrum. At moderate values of $B$ the spectra demonstrate the property of *combined invariance* (CI), i.e., invariance regarding simultaneous 90-degrees rotation of the sample and reversal of the polarization configuration. The CI was observed earlier for the spectra of spin-flip Raman scattering in undoped QWs[22] but was never reported for the PL spectra. At higher values of $B$, the CI becomes violated. We discuss possible reasons for that.

**2. Experimental.**

The PL spectra were taken in backward geometry, always in crossed linear polarizations of the polarizer and the analyzer. The PL was dispersed in a triple spectrometer and recorded using the CCD detector. Excitation came from a dye laser and, for the spectra presented here, was tuned to the energy ~30 meV higher than the QW PL. The sample was immersed in liquid helium at $T = 1.5$ K. The magnetic field up to 4.5 T was induced by a split-coil horizontal magnet and was perpendicular to the optical axis of the experiment (Voigt configuration).

The sample comprised a single 12 nm (Cd,Mn)Te QW sandwiched between $Cd_{0.85}Mg_{0.15}Te$ barriers and grown on a (001)-oriented substrate. The content of the magnetic $Mn^{2+}$ ions in the QW layer was 0.79% of the cation sites. The QW contained a two-dimensional electron gas with a reported[23] concentration $n_e = 2.1 \cdot 10^{11}$ cm$^{-2}$. Some previous experimental data regarding our sample can be found in Ref.24 (sample #405A). In-plane rotations of the sample were performed through warming up to the room temperature and re-positioning on the sample holder.

## 3. Results and discussion.

The PL spectra of QWs with a 2D electron gas at this (intermediate) level of electron concentrations typically include several spectral features. These features, or contours, are close in energy and broadened, so that they noticeably overlap. In addition, the QW layer in our sample is formed by a diluted magnetic semiconductor (DMS), a material characterized by a giant spin-dependent transformation of the band structure in the applied magnetic field.[25] The combination of these two factors makes the shape of the PL spectra both field- and polarization-dependent and rather individual.

Fig.1 illustrates the *B*-field evolution of the PL collected in two opposite linear polarizations (panels (a) and (b), respectively). Here and below, the experimental configurations are specified by three-letter indices as follows. The first letter specifies the magnetic field orientation in the laboratory reference frame; this field was always horizontal (*H*). The second letter specifies the orientation of the [110]-axis of the sample which can be either horizontal (*H*) or vertical (*V*). The third letter specifies the orientation of the analyzer which can be either horizontal (*H*) or vertical (*V*) as well.[26] All in all, four physically nonequivalent configurations result: *HHH*, *HHV*, *HVH* and *HVV*.

The PL spectra in both configurations presented in Fig.1 are mainly formed by three features. The L1 feature goes down in energy as the field is applied, thus it is naturally perceived as a lower (downshifting) Zeeman branch of some state. Two branches of the L2 feature represent the upper and the lower Zeeman partners of another state.[24] As shown in Ref.24, the energetic separation of the split branches of the L2 line as a function of *B* follows the spin splitting of the conduction-band electron.

The interpretation of the PL spectra like those in Fig.1, either in terms of the Fermi Sea or in terms of multi-particle states, has been discussed elsewhere.[23,24,27] That discussion is not relevant to our present subject. Here we just note that in Fig.1, spectra recorded in opposite polarizations are obviously different. This is very clearly visible in Fig.2: the use of opposite orientation of the analyzer leads to a clearly different spectrum. Moreover,

Fig.2 shows that any 90-degrees rotation of the sample leads to a strikingly different spectrum as well. The latter fact confirms the in-plane anisotropy of the sample and non-equivalence of its (nominally equivalent) [110] and [1$\bar{1}$0] directions.

Remarkably, if one combines the 90-degrees rotation of the sample with the reversal of the laser polarization and detected polarization (i.e., performs a *combined reversal* operation, CR), the resulting PL spectrum turns out to be very similar to the initial one (Fig.2). This is the CI property.[22]

As established in Ref.22, the CI in quantum wells or quantum dots originates from the pseudo-isotropic spin structure of the valence-band ground state. In turn, the *g*-factor of the ground state holes is induced by an in-plane uniaxial perturbation (e.g., deformation) which mixes the heavy-hole and the light-hole subbands. The physics of the pseudo-isotropic *g*-factor is well understood. In brief, the in-plane *g*-factor of the heavy hole is a block 2x2 of zeroes. An in-plane perturbation admixes the light-hole states and induces a non-zero *g*-factor whose symmetry fully reproduces that of the perturbation. As a result, the Zeeman splitting of the hole states is linear in *B* and does not depend on the orientation of the field, but the polarization selection rules are different as compared to the case of the isotropic g-factor. The (linear) polarization of every single optical transition between the Zeeman-split states follows the orientation of the crystal rather than that of the magnetic field. In particular, these selection rules result in the CI.

The QW sample studied here presents a convenient model system for demonstration of the CI: The spectra are individual enough and undergo pronounced polarization-dependent changes in the magnetic field. This has allowed the first demonstration of the CI by means of PL spectra. In addition, no examples of the pseudo-isotropic behavior of the valence-band *g*-factor were presented earlier for QWs containing electron gases. Our results show that this physics safely works in doped QWs too.

The CI is maintained as long as the electron spin splitting is controlled by the isotropic g-factor while the hole spin splitting is governed by the pseudo-isotropic g-factor. In stronger magnetic fields, the CI may become broken because of the *B*-superlinear contributions to the hole spin splitting. In fact, this violation of the CI could be observed in our sample (Fig.3). One can see that for two experimental configurations related to each other by the CR operation, the PL features reveal similar behavior in weak fields but deviate from each other in the fields above ~1.5 T, where apparently the $B^3$ contribution to the hole spin splitting[28] becomes comparable to the linear term.

## 4. Conclusions.

By means of the PL spectra of a DMS quantum well doped by electrons, we demonstrated the characteristic behavior of polarization selection rules for optical transitions. Typically for epitaxial quantum wells grown of cubic semiconductors along the [001] direction, the true symmetry turns out to be lower than the nominal symmetry ($D_{2d}$), with [110] and [1$\bar{1}$0] directions being non-equivalent to each other. This leads to a sensitivity of the PL spectra regarding the respective orientation of the sample and the magnetic field. Remarkably, the CI behavior observed in this sample below ~1.5 T confirms that the hole spin splitting (which cannot be spectrally resolved) is linear in *B* and has been induced by an in-plane distortion of the QW layer, rather than by an intrinsic physical reason. As the field was further increased, a violation of the CI showed up, manifesting the onset of the superlinear contributions to the spin splitting of hole states.

Conclusions of the present report have little relation to any microscopic interpretation of the PL spectra of the system under study. They were derived on the basis of simple symmetry operations over the sample and the polarization optics. The quite distinctive PL spectrum of our system and its pronounced polarization-dependent behavior in the external magnetic field allowed the first observation of the CI behavior in the photoluminescence, and also, for a quantum well system containing a 2D electron gas.

**5. Acknowledgements.**

This work was partially supported by the Russian Ministry of Science and Education (contract No. 11.G34.31.0067), by SPbSU (grant No. 11.38.213.2014), by RFBR (projects 13-02-00316, 15-52-12019) and by the Polish National Science Center (grant No. 2014/14/M/ST3/00484). AK gratefully acknowledges support from Dmitry Zimin "Dynasty" Foundation.

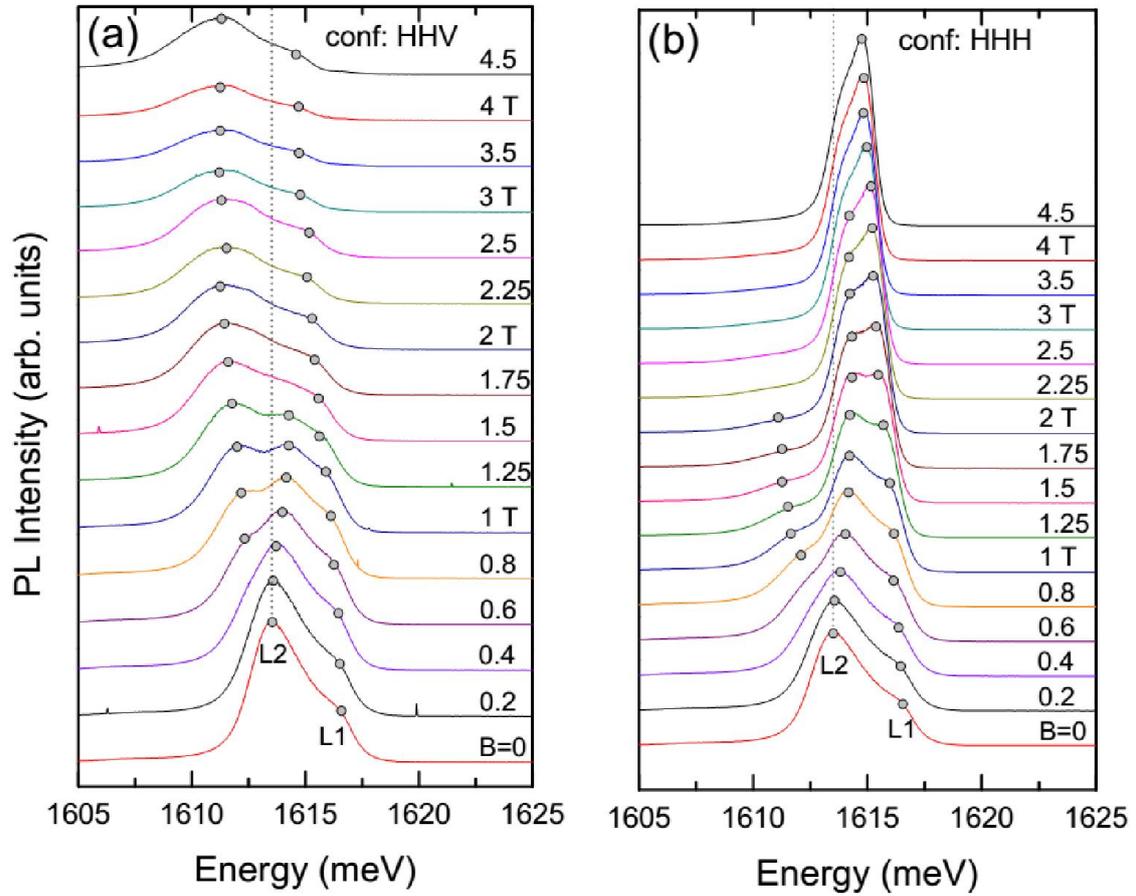

Fig. 1. Photoluminescence spectra of the n-type (001)-$Cd_{0.99}Mn_{0.01}$ QW taken at different values of the in-plane magnetic field (specified at the curves) and at two different orientations of the analyzer: perpendicular to (a) and along (b) the magnetic field direction. The main features forming the spectra are marked by grey circles.

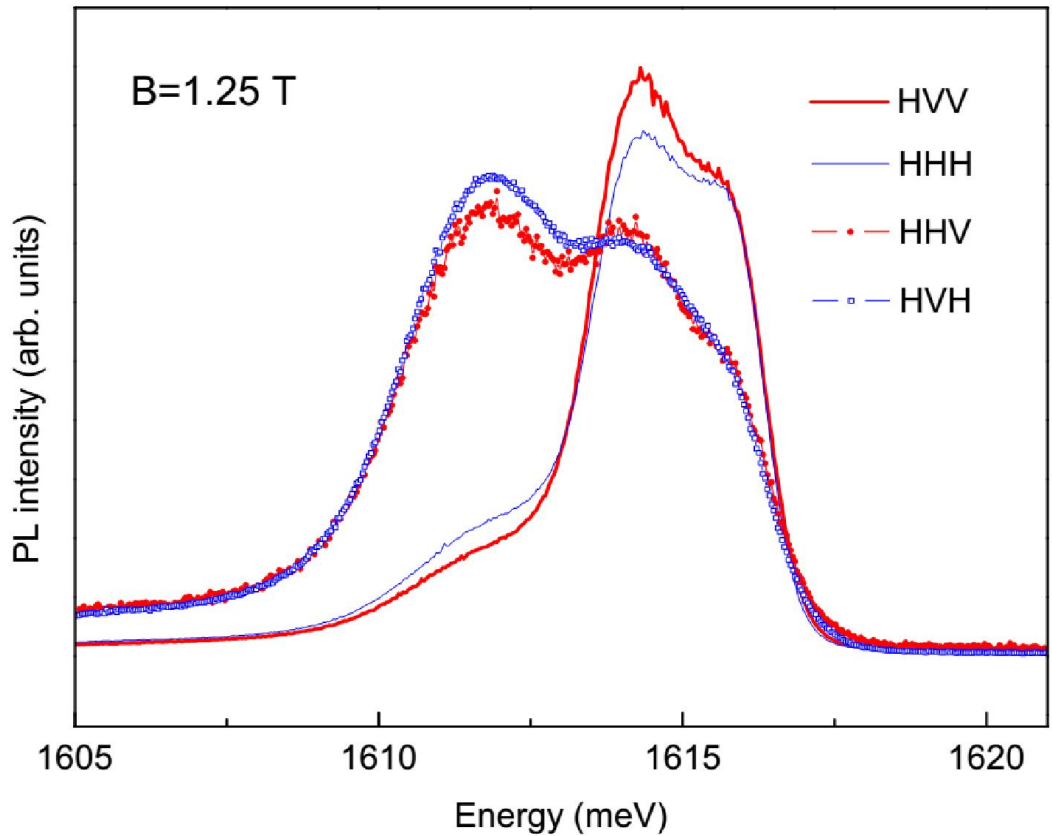

Fig. 2. The CI property demonstrated by the PL spectra taken at the magnetic field $B = 1.25$ T and four different configurations of the experiment. The spectra differing by the sample orientation only (within the pairs *HVV*, *HHV* and *HHH*, *HVH*) are clearly different, proving the in-plane anisotropy of the sample. The spectra differing in the orientation of the analyzer only (pairs *HVV*, *HVH* and *HHH*, *HHV*) are different too. But simultaneous rotations of the sample and the analyzer result in remarkably like spectra (pairs *HVV*, *HHH* and *HHV*, *HVH*).

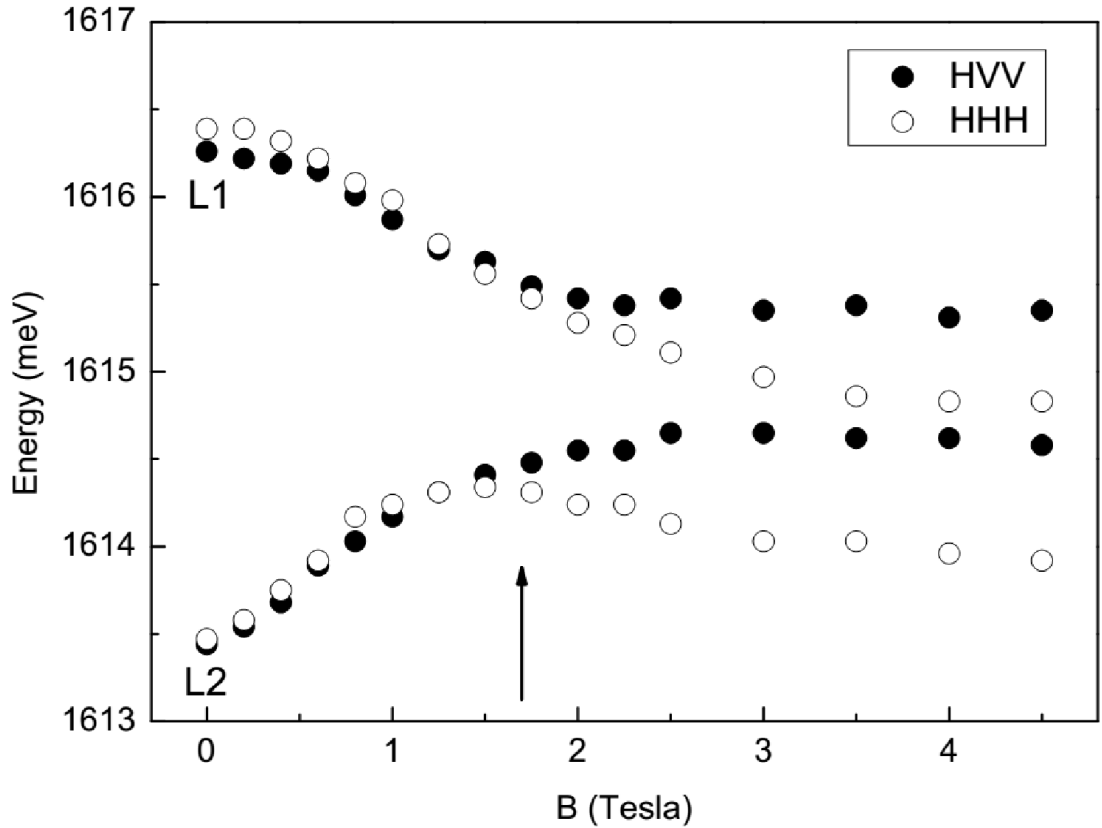

Fig.3. Violation of the CI at strong magnetic fields. Points show peak positions of the lower Zeeman branch of the L1 line and of the upper Zeeman branch of the L2 line (see Fig.1) versus the magnetic field strength. Open and closed circles correspond to two different experimental configurations which are related to each other by the CR operation. The peak positions coincide at weak fields (in compliance with the CI) but deviate above ~1.5 T, where the *B*-nonlinear contributions may get control over the hole spin splitting.